\documentstyle[aps,epsf,rotate,multicol]{revtex}

\begin{document}

\draft

\title{Small-world networks and the conformation space \\ of 
a lattice polymer chain}

\author{Antonio Scala, Lu\'{\i}s A. Nunes Amaral, and Marc
Barth\'el\'emy}

\address{Center for Polymer Studies and Department of Physics, 
	Boston University, MA 02215}


\maketitle

\begin{abstract}

We map the conformation space of a simple lattice polymer chain
to a network, where (i) the vertices of the network have a one-to-one
correspondence to the conformations of the chain, and (ii) a link
between two vertices indicates the possibility of switching from one
conformation to the other by a single Monte Carlo move of the chain.
We find that the geometric properties of this network are similar to
those of small-world networks, namely, the diameter of conformation
space increases, for large networks, as the logarithm of the number of
conformations, while locally the network appears to have low
dimensionality.

\end{abstract}

\begin{multicols}{2}


The physical properties of polymers are the focus of a lot of
attention due to (i) their role in many new advanced materials with
important technologic applications \cite{techpolymers}, and (ii) their
role in biological processes \cite{biopolymers}.  An especially
important unsolved problem in polymer studies is protein folding.
Solving this problem has recently become more urgent as several
studies suggest that a number of human diseases, including
Parkinson's, Alzheimer's and British dementia, may be due to the
aggregation of misfolded proteins\cite{Lansbury99,Eaton99}.

The kinetics of protein folding are controlled by the structure of the
free-energy landscape\cite{Wolynes}. Theoretical calculations predict
that in some cases the barriers in the free-energy landscape are quite
small\cite{ThirumalaiPortman}. Hence, the diffusion of a protein's
conformation on its energy landscape may be determined mainly by the
structure and connectivity of conformation space.  For this reason,
much work has been done on the structure of conformation space and
many models have been proposed, including tree structures, random
networks, and ultrametric spaces\cite{OgielskiVelikson}.  In spite of
all this work, very little is known about the conformation space of
polymers in general and proteins in particular.

Here, we present evidence supporting the hypothesis that the
conformation space of a lattice homopolymer chain may be a small-world
network.  These networks ---which appear as the result of randomly
replacing a fraction $p$ of the links of a $d$-dimensional lattice
with new random links\cite{Watts}--- interpolate between the two
limiting cases of a regular lattice ($p=0$) and a random graph
($p=1$).  The small-world regime is characterized by the properties
(i) that a local neighborhood is preserved ---as for regular
lattices\cite{Watts}--- and (ii) that the average shortest distance
between two vertices increases logarithmically with the number of
vertices $n$ of the network ---as for random graphs\cite{Bollobas85}.

\begin{figure}
\narrowtext
\centerline{
\epsfysize=0.8\columnwidth{\epsfbox{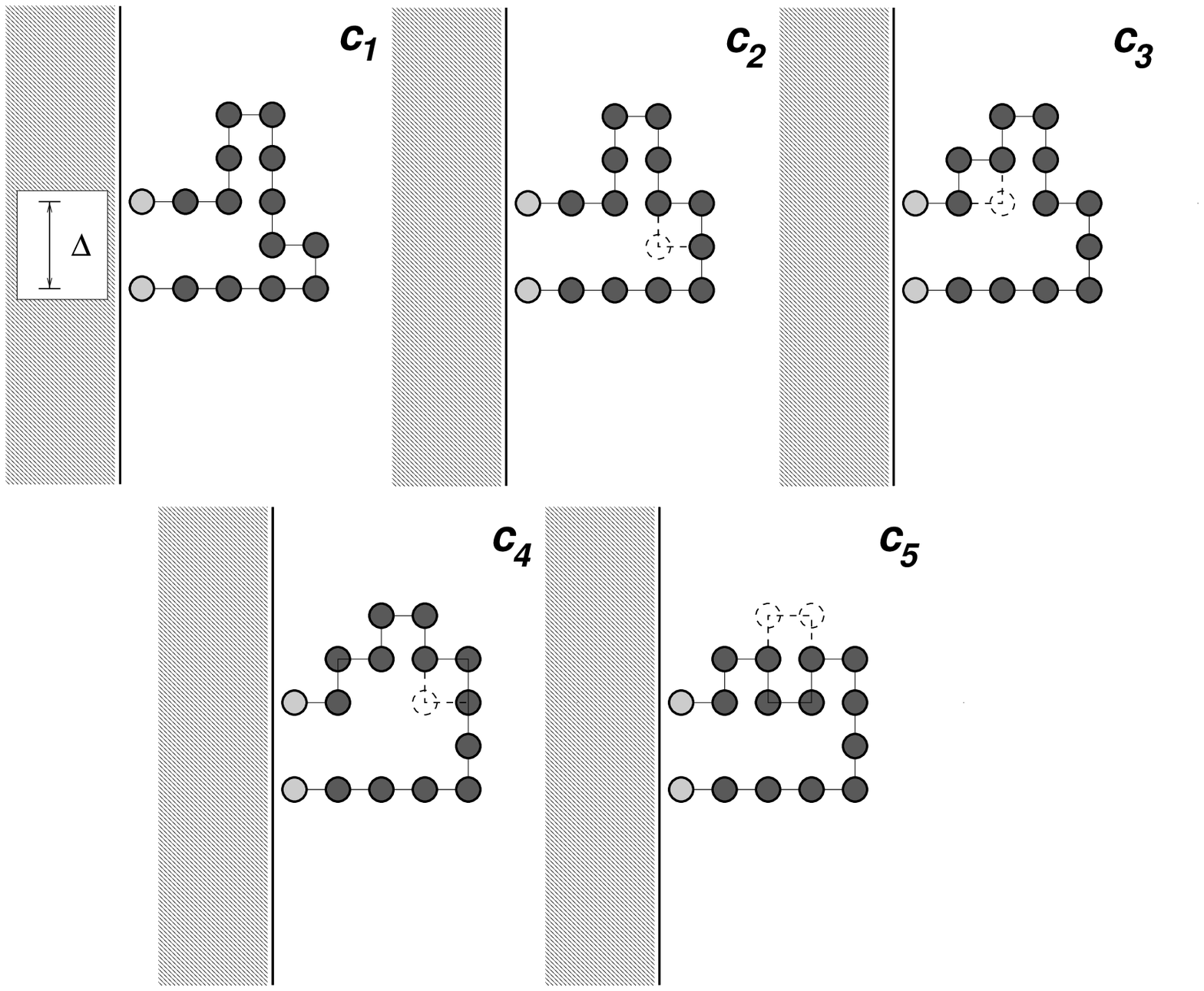}} 
}
\centerline{
\epsfysize=0.6\columnwidth{\epsfbox{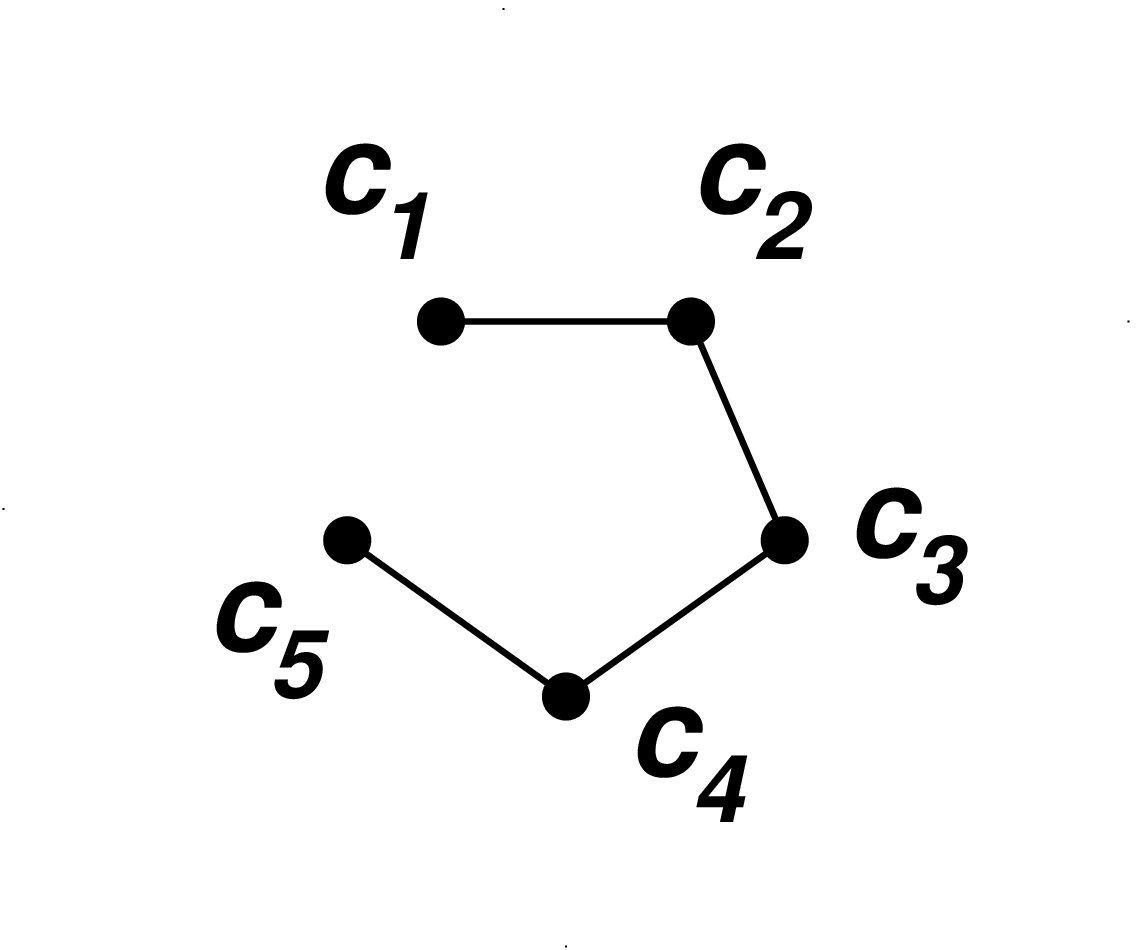}}
}
\caption{ The conformations of a lattice polymer chain and the mapping
of the conformation space to a network.  {\bf (a)} The five panels
labeled $c_1$ to $c_5$ show 5 different conformations of a 15-mer
constrained to have the end monomers fixed a distance $\Delta$ apart.
We use circles with a light shade to represent the fixed monomers, and
dark circles to represent the moving monomers. For numerical
convenience, we constrain the chain to occupy the half plane bounded
by the thick black line.  Note that the same results are obtained for
other boundary conditions.  The conformation of the chain evolves by
the usual Monte Carlo elementary moves: the corner flip and the
``crankshaft''.  For example, the chain can switch from conformations
$c_1$ to $c_2$, $c_2$ to $c_3$, and $c_3$ to $c_4$ by single corner
flips.  In panels $c_2$ to $c_5$, we use dashed lines and dashed
circles to represent the position of the monomer moved from the
previous conformation.  We use conformations $c_4$ and $c_5$ to
illustrate the ``crankshaft'' move, which involves the simultaneous
movement of two monomers.  {\bf (b)} Mapping of the conformation space
of a chain to a network.  We first allocate a vertex of the network to
each allowed conformation of the chain.  We then create a link between
two vertices if the two corresponding conformations differ by a single
elementary Monte Carlo move.  For example, since we can switch between
conformations $c_1$ and $c_2$ by a single move, we put a link between
the vertices $c_1$ and $c_2$.  On the other hand, because $c_5$ cannot
be reached from $c_1$ through a single elementary move, we do not
place a link between the corresponding vertices.  }
\label{f.chain}
\end{figure}

To gain some insight on the conformation space of real polymers, we
consider here the conformation space of a 2-dimensional lattice
polymer chain.  To study the geometrical properties of conformation
space, we map it onto a network~\cite{Du99a}.  We first enumerate {\it
all\/} allowed conformations of the chain.  We then identify (i) each
conformation of the chain with a vertex of the network, and (ii) the
possibility of changing from one conformation to another, through a
{\it single\/} Monte Carlo move of the chain, with the existence of a
link between the corresponding vertices (Fig.~\ref{f.chain}).

Our study relies on two important simplifications.  The first
simplification is to use a lattice model.  On this regard, note that
due to the limited number of allowed equilibrium angles between two
monomers, the modeling of a polymer by a linear chain on a lattice may
be experimentally justified\cite{Chan-Klimov}.  In fact, even for the
study of proteins, lattice models led to important insights
\cite{Du99a,Chan-Klimov}.  The second simplification is to neglect
interactions.  On this regard, note that here we are only interested
in determining the {\it geometric and structural\/} properties of
conformation space, and that interactions affect the {\it rate of
transfer\/} between the allowed conformations
\cite{Eaton99,Du99a,HagenSabelko}.  Moreover, our results will also
apply to any polymer chain in the limit of very high temperature for
which the monomer interactions are mostly due to steric effects
\cite{Pitard98}.


We first calculate the geometric properties of conformation space for
the simplified lattice polymer model and compare these properties with
the predictions of several models for conformation space.  This
approach is {\it different\/} from the ones considered so far in the
literature.  We do not start by postulating a particular type of
structure with some theoretically-desirable properties but, instead,
try to generate the full network describing conformation space and
compare its properties with different models.

\begin{figure}
\narrowtext
\centerline{
\epsfysize=0.7\columnwidth{\epsfbox{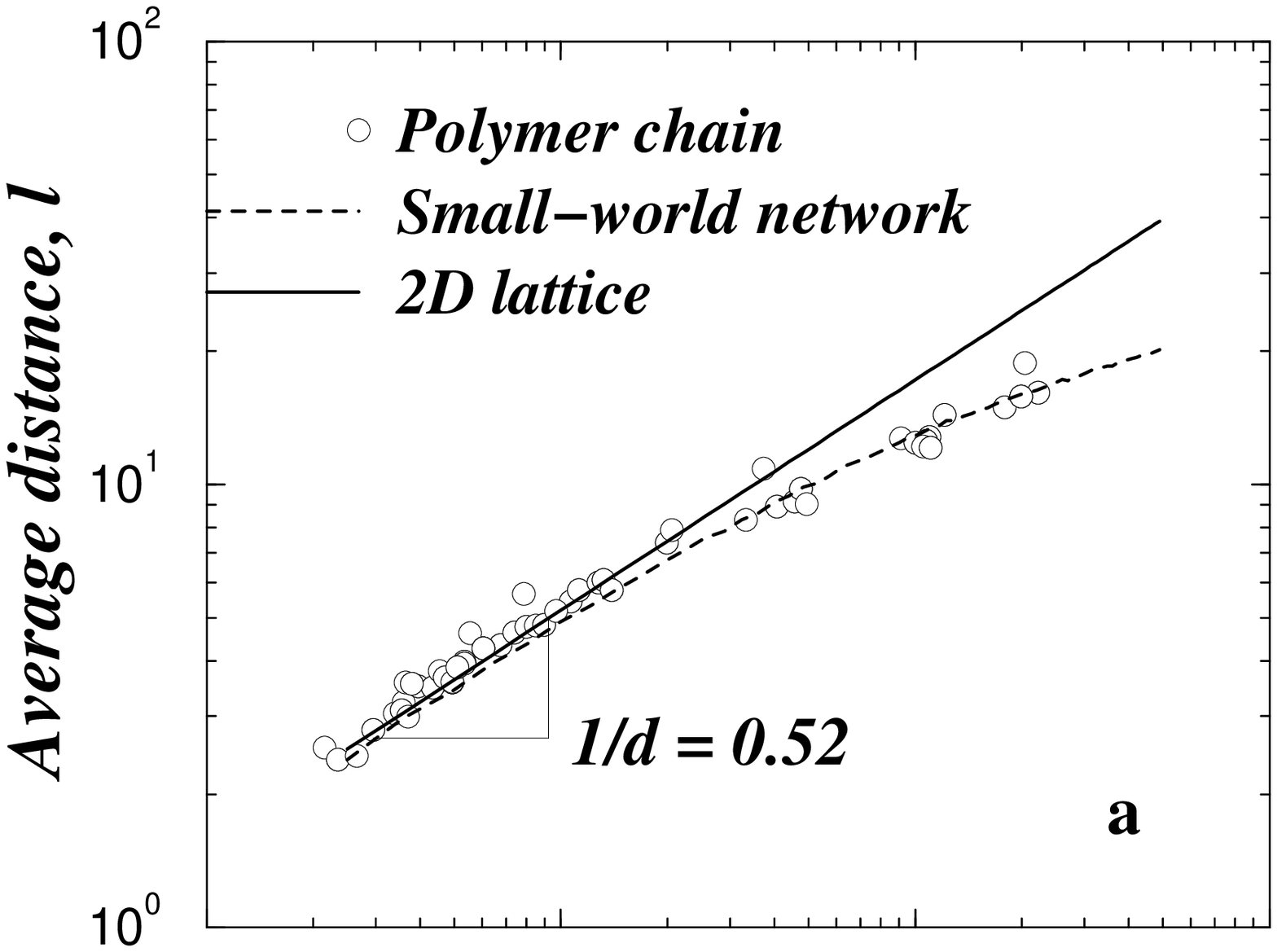}}
}
\vspace*{-0.9cm}
\centerline{
\epsfysize=0.7\columnwidth{\epsfbox{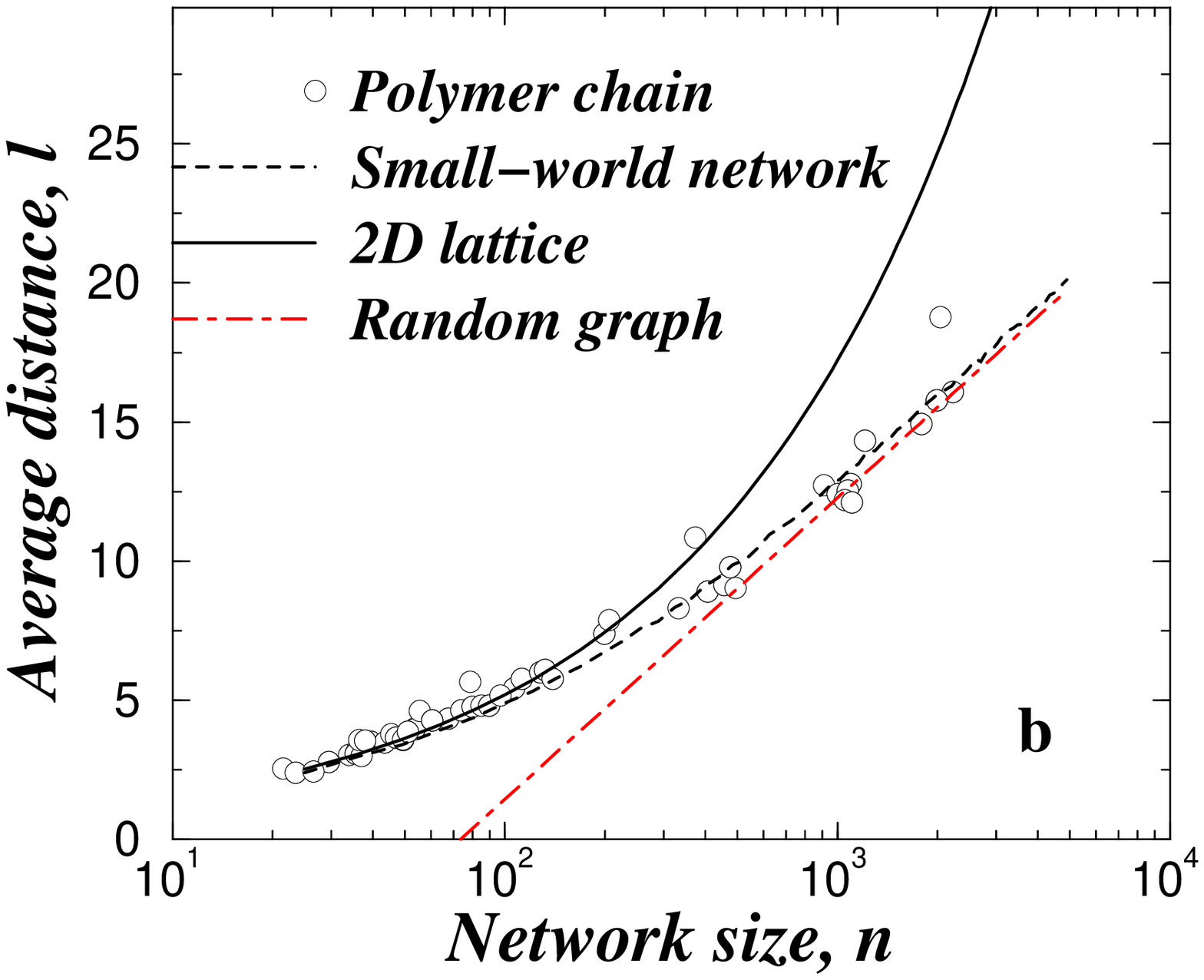}}
}
\caption{ Average shortest distance between chain conformations. We
generate all the possible conformations of the polymer chain defined
in Fig.~\protect\ref{f.chain} for $\Delta = 1, \dots, 6$ and for
chains with $m=6,\dots, 15$ monomers.  For each pair of values
$(\Delta, m)$, we identify in the conformation space all the different
{\it connected networks\/} and calculate their size $n$.  Here, we
calculate the average shortest distance between every pair of vertices
using to the breadth-first search algorithm.  {\bf (a)} Loglog plot of
the average distance, rescaled according to the value of average
connectivity\protect\cite{Watts}.  For $n < 100$, the average distance
$\ell$ increases as $n^{1/2}$, as it would for a 2-dimensional
lattice.  {\bf (b)} Same data but in log-linear plot. For $n > 1000$,
we observe $\ell \sim \ln n$, as for a random network, and as
predicted by Eq.~(\protect\ref{e.ell}).  Our results are consistent
with the case of a small-world network with $p \approx 10^{-3}$,
clearly ruling out a low dimensional lattice as a model of
conformation space.  Note that for this and all other figures, the
results for the small-world network have been average over 100
realizations of the network while each symbol ``$\circ$'' corresponds
to a single conformation network. Hence, there is far less noise for
the small-world networks data. }
\label{f.length}
\end{figure}

As a first test, we study the dependence of the average shortest
distance $\ell$ between any two vertices in the network on the number
$n$ of vertices.  Note (i) that the size of the network equals the
number of allowed conformations of the chain, and (ii) that the
distance between two vertices equals the minimum number of elementary
moves of the chain necessary to switch between the two corresponding
conformations. For a tree structure or a random network we expect a
logarithmic increase of $\ell$ with $n$, while for a $d$-dimensional
lattice we expect an algebraic increase: $\ell \sim n^{1/d}$.  In
contrast, for a small-world network, $\ell$ follows the scaling
law\cite{Barthelemy99}
\begin{equation}
\ell(n,p) \sim (n^*)^{1/d} f( n / n^*)\,,
\label{e.ell}
\end{equation}
where the scaling function $f(u)$ has the limits $f(u) \sim u^{1/d}$
for $u\ll 1$ and $f(u) \sim \ln u$ for $u\gg 1$; $n^* \sim p^{-1}$ is
a crossover size that separates the large- and small-world regimes,
and $p$ is the fraction of ``rewired'' links\cite{Barthelemy99}.
Figure~\ref{f.length} displays our results for the polymer chain,
which suggest that the conformation space can be described by a
small-world network with $p \approx 10^{-3}$.  Figure~\ref{f.length}
clearly rules out the possibility that, for $n \gg 1$, the
conformation space is a low dimensionality lattice.

\begin{figure}
\narrowtext
\centerline{
\epsfysize=0.7\columnwidth{\epsfbox{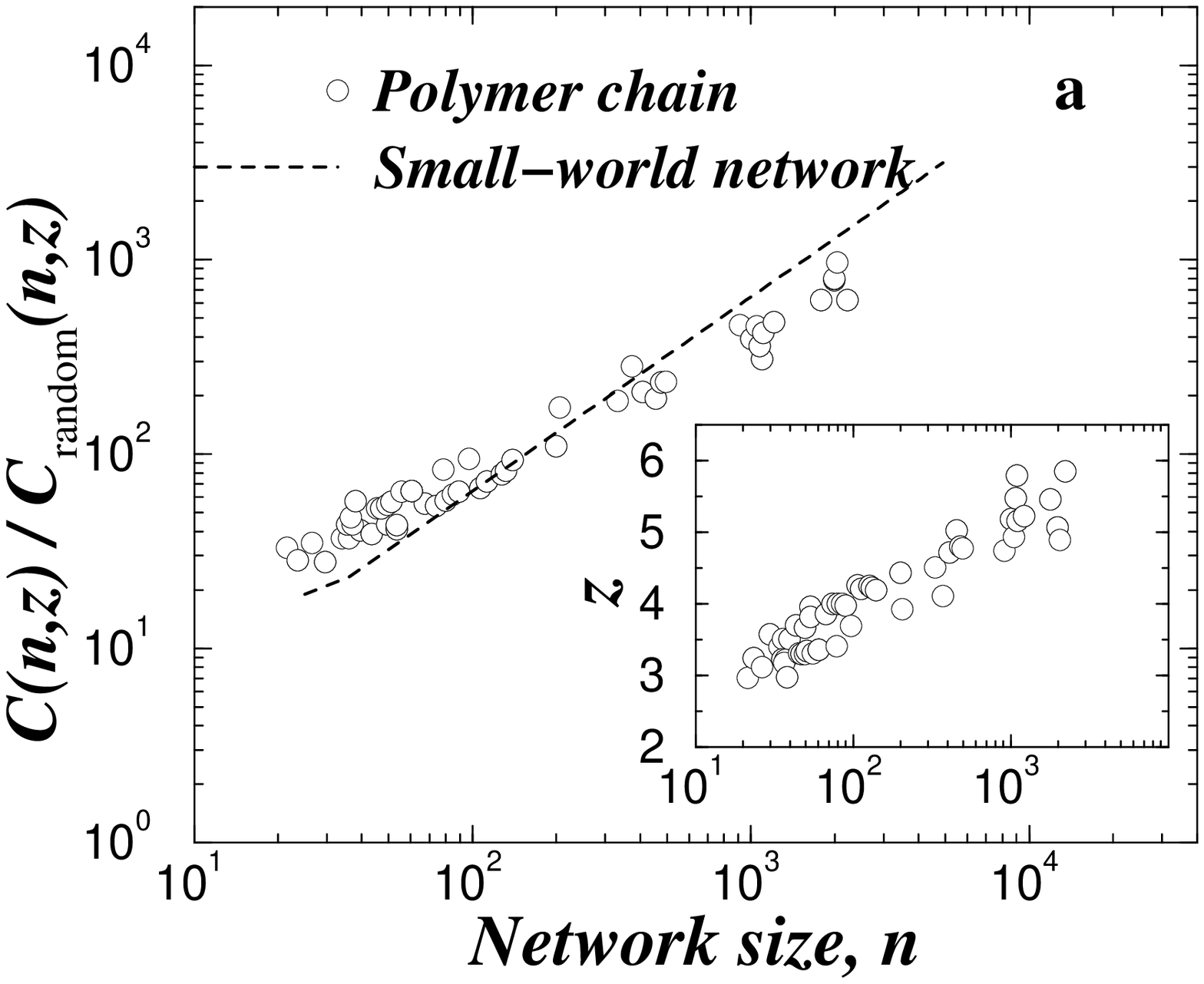}}
}
\vspace*{0.0cm}
\centerline{
\epsfysize=0.7\columnwidth{\epsfbox{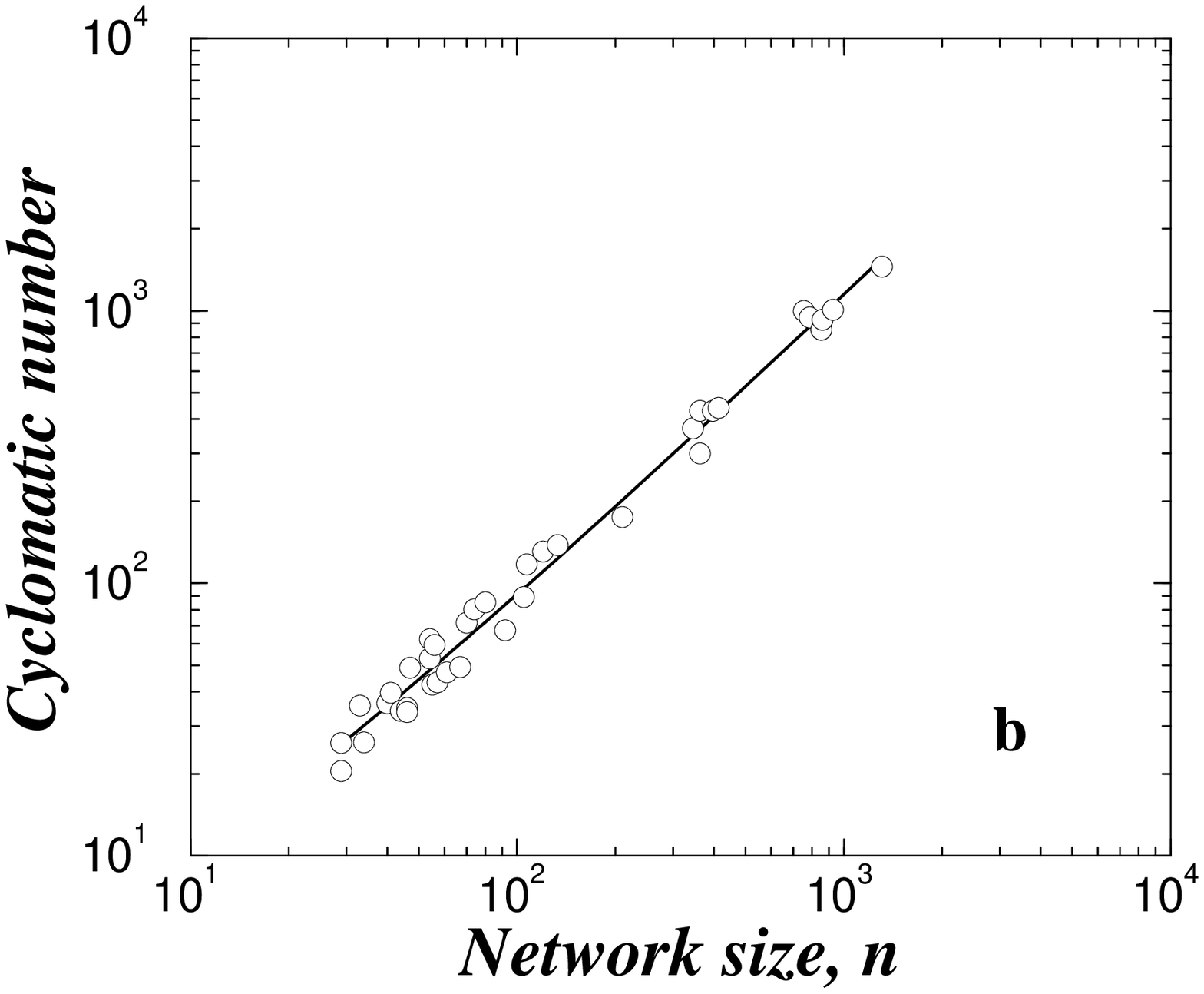}} 
}
\centerline{
\epsfysize=0.7\columnwidth{\epsfbox{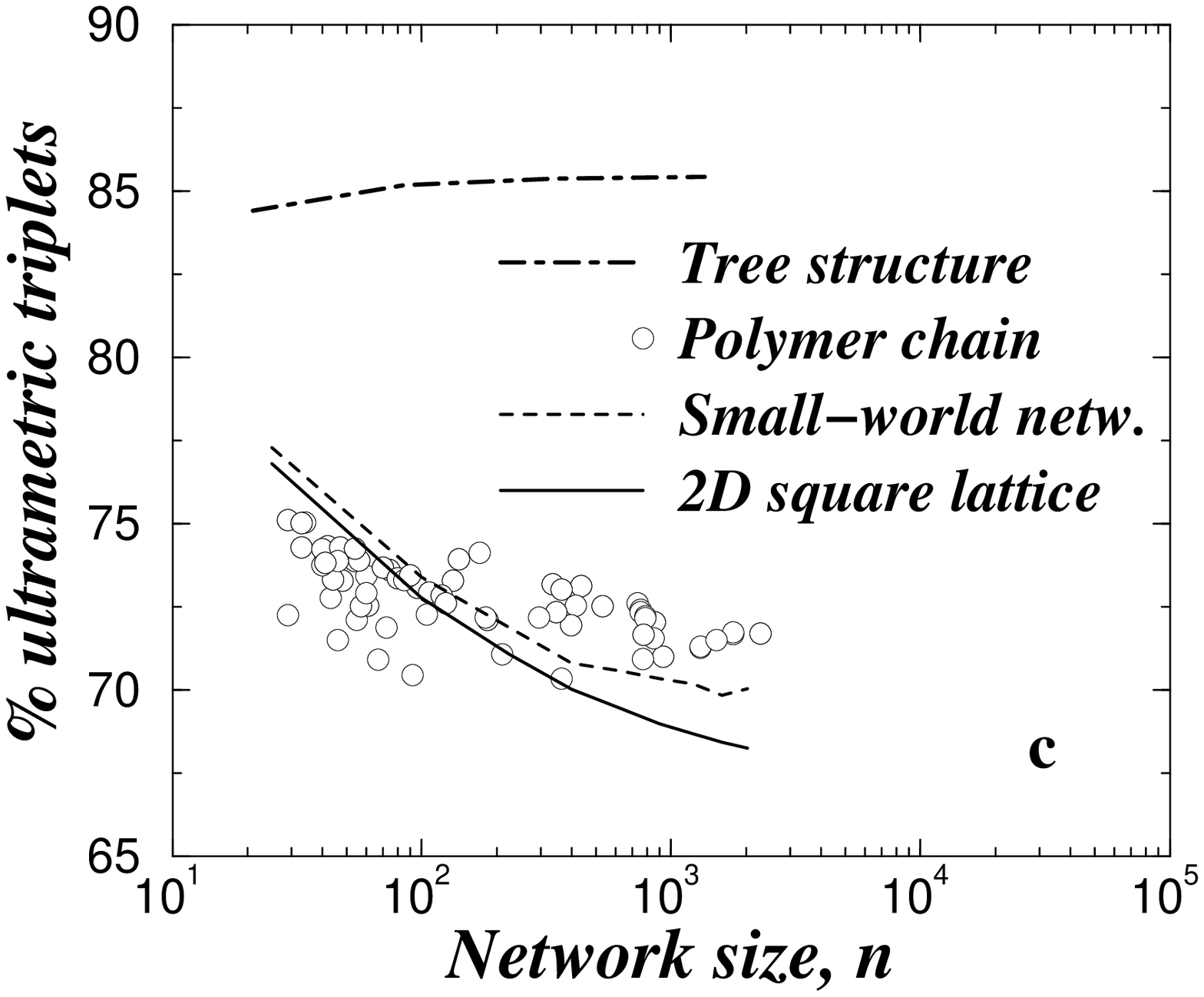}}
}
\caption{ Structure of conformation space. {\bf (a)} Normalized
clustering coefficients for the polymer chain, $C(n,z) / C_{\rm
random}(n,z) \simeq n C(n,z) / z$, where $z$ is the average
connectivity. For a small-world network, the clustering coefficient
will be approximately a constant (for fixed $p$), and the normalized
clustering coefficient increases linearly with $n$.  Our results
clearly rules out a purely random structure for conformation space
since the normalized clustering coefficient for the polymer chain
conformation space is orders of magnitude larger than the value for a
random network. Note that the apparently logarithmic discrepancy
between the two curves in the figure may be due to the logarithmic
increase of $z$ with $n$, which is shown in the inset.  {\bf (b)}
Cyclomatic number. The cyclomatic number for a tree structure is
identically zero because there are no loops.  For a generic
network\protect\cite{Berge76}, the cyclomatic number increases as
$nz$.  Hence, we plot the cyclomatic number for the polymer
conformation space normalized by the connectivity $z$.  As expected,
we observe a linear increase with size $n$.  {\bf (c)} Percentage of
ultrametric triplets.  We calculate the number of triplets obeying the
relation $d_{AC} \le \max(d_{AB}, d_{BC})$. For a purely ultrametric
space, all triplets have distances obeying this relation.  For
2-dimensional square lattices and for small-world networks only
slightly more than 2/3 of the triplets ---which is the lower bound---
obey the ultrametric relation.  In contrast, for random networks and
tree structures, we find a higher percentage of triplets to be
ultrametric. It is apparent that values for the random network and
tree structure are not consistent with the measured values for the
polymer conformation space.  Note that each data point for the polymer
chain corresponds to a single conformation network while the results
for the small-world network represent an average over 100 networks. }
\label{f.structure}
\end{figure}

As a second test, we study the local structure of conformation space
and compare it with that of a random network\cite{Watts}.  To this
end, we calculate the clustering coefficient $C$, which is defined as
the average ratio of the number of existing links between neighbors of
a vertex and the maximum number of possible links.  For a random
network\cite{Watts}, we expect $C \simeq z/n$, where $z$ is the
average connectivity of the network. In contrast, small-world networks
have values of $C$ of the same order of magnitude as those of regular
lattices, because only a small percentage of links are different from
those in the lattice\cite{Watts}.  In
Fig.~\ref{f.structure}(a), we compare the values of $C$ obtained for
the networks with the values of $C$ for random networks with the same
size and connectivity. Clearly, the measured clustering coefficients
are much larger than the expected values for random networks, ruling
out a purely random structure for conformation space.

As a third test, we calculate the number of elementary loops in
conformation space ---usually referred to as the cyclomatic
number\cite{Berge76}--- and compare the results with those for a tree
structure.  For a tree structure the cyclomatic number is identically
zero while for all other networks it increases linearly with $n$.
Figure~\ref{f.structure}(b) shows that the cyclomatic number for the
polymer conformation space clearly increases with $n$, consistent with
a small-world network but ruling out a tree structure for conformation
space.

As a final test, we calculate the percentage of triplets $\{A,B,C\}$
of vertices in conformation space whose distances obey an ultrametric
relation\cite{OgielskiVelikson}: $d_{AC} \le \max(d_{AB}, d_{BC})$.
Figure~\ref{f.structure}(c) shows the percentage of ultrametric
triplets for conformation space.  The percentage of ultrametric
triplets is significantly smaller than 100\%, ruling out a purely
ultrametric structure for conformation space.  Moreover, the measured
percentage of ultrametric triplets also rules out the random network
and the tree structure as descriptions of conformation space.

In summary, the regular lattice is rejected by the first test, the
tree structure is rejected by the first, third, and fourth tests, and
the random network is rejected by the first, second and fourth tests.
Hence, we conclude that the geometrical properties of the conformation
space of a lattice polymer chain are consistent with those of
small-world networks but not with the geometric properties of the
other geometries discussed in the literature.


Next, we address the implications of our finding that the conformation
space of a lattice polymer chain may be a small-world network.  A
central problem in protein folding is the characterization of
relaxation processes.  The kinetics of a protein's conformations can
be mapped to a random walk on conformation space.  Naturally, the
model studied here is too simplistic to enable us to understand
protein folding, however, the study of diffusion on conformation
spaces with small-world geometries may still develop our understanding
of relaxation in protein folding.  Hence, we consider next the problem
of diffusion on a small-world network.  The usual way to solve a
diffusion problem is to obtain the density of states $\rho(\lambda)$
of the transition matrix of the system\cite{Bray88}.  For a
small-world network, one has\cite{Bray88,Monasson99}
\begin{equation}
\rho(\lambda) \sim \lambda^{d/2 - 1}~\exp(-p / \sqrt{\lambda}) \,,
\label{e.density}
\end{equation}
where the $\lambda$'s are the eigenvalues of the transition matrix for
the network.  To quantify the relaxation properties, we calculate the
probability of return to the origin, usually considered in the study
of random walks\cite{Havlin87}, which is the Laplace transform of
(\ref{e.density}),
\begin{equation}
P_0(t) = \int_{1/\ell^2}^{\infty} d\lambda~\rho(\lambda)~\exp( 
-\lambda t) \,,
\end{equation}
where the lower limit in the integral is a cut-off that takes into
account the finite size of the network.  The return probability has
the scaling behavior
\begin{equation}
P_0(t) - P_0(\infty) \sim \left\{
\begin{array}{ll}
      \mbox{$t^{-d/2}$}             & \mbox{$t \ll t_1$}          \\
      \mbox{$\exp(-(p^2 t)^{1/3})$} & \mbox{$t_1 \ll t \ll t_2$}  \\
      \mbox{$\exp(-t/{\ell}^2)$}    & \mbox{$t \gg t_2$}          \\
\end{array} \right. ,
\label{e.P0}
\end{equation}
where $P_0(\infty)=1/n$ and the two crossover times scales can be
written, in the small-world regime, as
\begin{equation}
\left\{
\begin{array}{l}
t_1 \sim 1/p^2 \sim (n^*)^2 \\
t_2 \sim p{\ell}^3 \sim (n^*)^2 (\ln n)^3 \\
\end{array} \right. ,
\label{e.times}
\end{equation}
where we used the result $\ell \sim n^* \ln n$ \cite{Barthelemy99}.
The results (\ref{e.density}-\ref{e.times}) are consistent with our
numerical simulations both for small-world networks and for the
conformation space of the lattice polymer chain.


The description of the conformation space as a small-world network may
be useful for other complex disordered systems such as spin-glasses
\cite{Wales-Franzese}.  Specifically, small-world networks combine the
features of apparent {\it infinite dimensionality\/} and low
connectivity thought to be important for glassy relaxation
\cite{Wales-Franzese,Campbell94}.  It has been hypothesized, that in
order to reproduce stretched exponential decays
\cite{HagenSabelko,LimOnuchic}, the space of accessible conformations
must have a tree-like structure\cite{Bray88,Campbell94}.  We have
shown that for conformation spaces with a small-world structure
stretched exponential relaxation can appear as an intrinsic feature of
the space's geometry.





\end{multicols} 

\end{document}